\def\year{2020}\relax
\documentclass[letterpaper]{article} % DO NOT CHANGE THIS
\usepackage{aaai20}  % DO NOT CHANGE THIS
\usepackage{times}  % DO NOT CHANGE THIS
\usepackage{helvet} % DO NOT CHANGE THIS
\usepackage{courier}  % DO NOT CHANGE THIS
\usepackage[hyphens]{url}  % DO NOT CHANGE THIS
\usepackage{graphicx} % DO NOT CHANGE THIS
\usepackage[mathscr]{euscript}
\usepackage{amsmath}
\usepackage{graphicx}  % DO NOT CHANGE THIS
\usepackage{amssymb}
\usepackage{algorithmic}
\usepackage{graphicx}
\usepackage{subcaption}
\title{Zoom in to where it matters: a hierarchical graph based model for mammogram analysis }
\author{Hao Du, Jiashi Feng, Mengling Feng}
\begin{document}

\maketitle
\begin{abstract}
%Accurate breast cancer detection in mammograms is challenging becuase the lesions are too tiny compared to full field digital mammograms (FFDMs) in breast cancer sreening. In clincal practice, radiologists review FFDMs using high resolution monitors and zoom in for close-ups for region of interests.
In clinical practice, human radiologists actually review medical images with high resolution monitors and zoom into region of interests (ROIs) for a close-up examination.
Inspired by this observation, we propose a hierarchical graph neural network to detect abnormal lesions from medical images by automatically zooming into ROIs.
We focus on mammogram analysis for breast cancer diagnosis for this study.
 Our proposed network consist of two graph attention networks performing two tasks: (1) node classification to predict whether to zoom into next level; (2) graph classification to classify whether a mammogram is normal/benign or malignant. The model is trained and evaluated on INbreast dataset and we obtain comparable AUC with state-of-the-art methods. 
\end{abstract}

\section{Introduction}

Recently, graph convolutional network (GCN) and its variants have generated considerable recent research interest in learning graph representations \cite{mamalet2012simplifying}. Applications of GCN and its variants have demonstrated new state-of-the-art results in various domains, such as applied chemistry, social network, citation network, computer vision and natural language processing \cite{kipf2016semi,liao2019lanczosnet,mamalet2012simplifying,chen2018fastgcn,yao2019graph}. Additionally, GCNs and their variants generated considerable recent research interest in medical imaging field. In 2018, Parisot et al employed GCNs to predict disease using a graph where nodes present individuals and features consist of both image and non-image data. GCNs are used as  a semi-supervised method to train on the labelled node and infer the labels of the unknown nodes, based on the graph structure and both image and non-image features \cite{parisot2018disease}. Shin et al. demonstrated combining both CNNs and GCNs to perform deep vessel segmentation \cite{shin2019deep}. CNNs are used to generate features and vessel probabilities while GCNs are employed to predict the presence of a vessel. By combining the outputs of both CNNs and GCNs, the model generates the final segmentation. 

We further observed that, in clinical practice, human radiologists actually review medical images with high resolution monitors and zoom into region of interests (ROIs) for a close-up examination.
Inspired by this observation, we propose a hierarchical graph neural network to detect abnormal lesions from medical images by automatically zooming into ROIs.
In this work, we focus on breast cancer detection in mammogram analysis. 

\begin{figure}[ht!]
	\centering
	\includegraphics[width=0.5\textwidth]{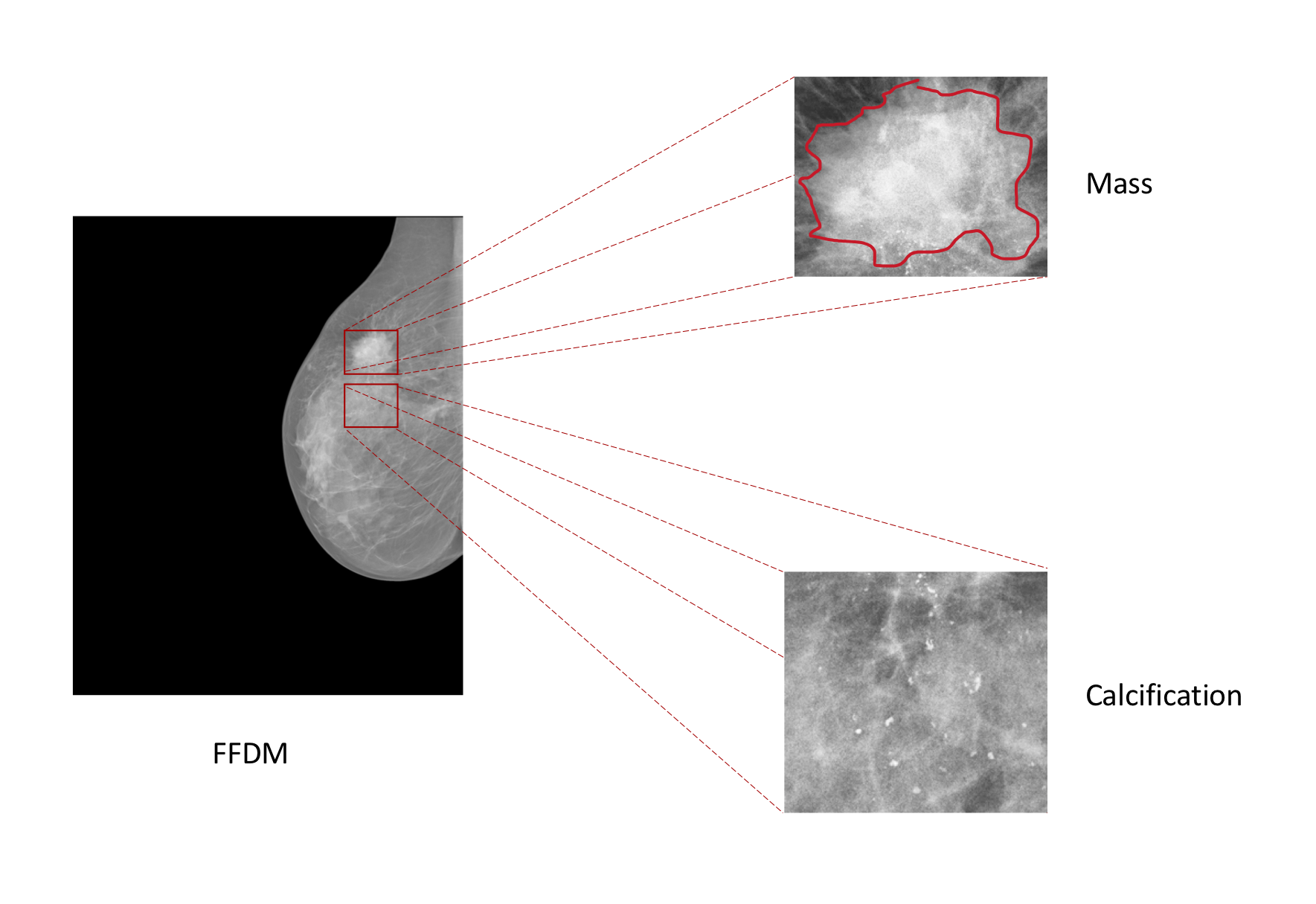}
	\caption{An example of FFDM image in the INbreast dataset. Two groups of lesions are identified by zooming into the specific patches on the right side. Mass is represented by red contours and microcalcifications are small, gray, rounded bright regions in the breast tissue. }
	\label{fig:1}
\end{figure}

Mammography is currently the most effective tool for breast cancer screening and early detection of the disease \cite{misra2010screening}. Breast screening for one patient includes the mammography images of two views for each breast: the craniocaudal (CC) view, which is a top to bottom view, and a mediolateral oblique (MLO) view, which is a side view. Currently, the examination of mammograms  mainly relies on human radiologists and facing the limitations of high false-positive rate, inconsistent interpretation accuracy and huge workload. \cite{noble2009computer,elmore1998ten,brodersen2013long,mccann2002impact}. Screening mammogram classification is challenging due to the difficulties in detection of lesions especially masses and microcalcifications. As demonstrated in Figure \ref{fig:1}, the average dimension of a full field digital mammogram (FFDM) is around 4000 $\times$ 3000. However, the dimensions of mass lesions are typically less than 100 $\times$ 100 \cite{shen2019deep} and the sizes of microcalcifications are at most 14 pixels \cite{zhang2019cascaded}. In clinical settings, the American College of Radiology recommends that all FFDMs should be viewed at their full acquisition resolution \cite{american2010practice}. Radiologists review the FFDM images using high resolution monitors and zoom in for close-ups of regions of interests (ROIs). 

Recently, deep learning methods have demonstrated big improvements on automated analysis of mammograms \cite{aboutalib2018deep,kim2018applying,hamidinekoo2018deep,burt2018deep,kooi2017large,agarwal2019automatic,ribli2018detecting,shen2019deep}. The challenges of lesion dimensions are addressed using established object detection and anomaly detection methods \cite{ribli2018detecting,zhang2019cascaded} or fully image classification methods on resized images with pre-trained models \cite{shen2019deep}. In this study, we propose a graph neural network (GNN) based method for end-to-end breast cancer detection in FFDM images. We use hierarchical graph based method to model the zoom-in mechanism of radiologists' operations. The proposed method is able to automatically zoom into the lesion ROIs and detect breast cancer based on overall graph hierarchy and specific ROIs in the mammograms.

\section{Methods}
\begin{figure*}[ht!]
	\centering
	\begin{subfigure}{\textwidth}
		\centering
		\includegraphics[width=\textwidth]{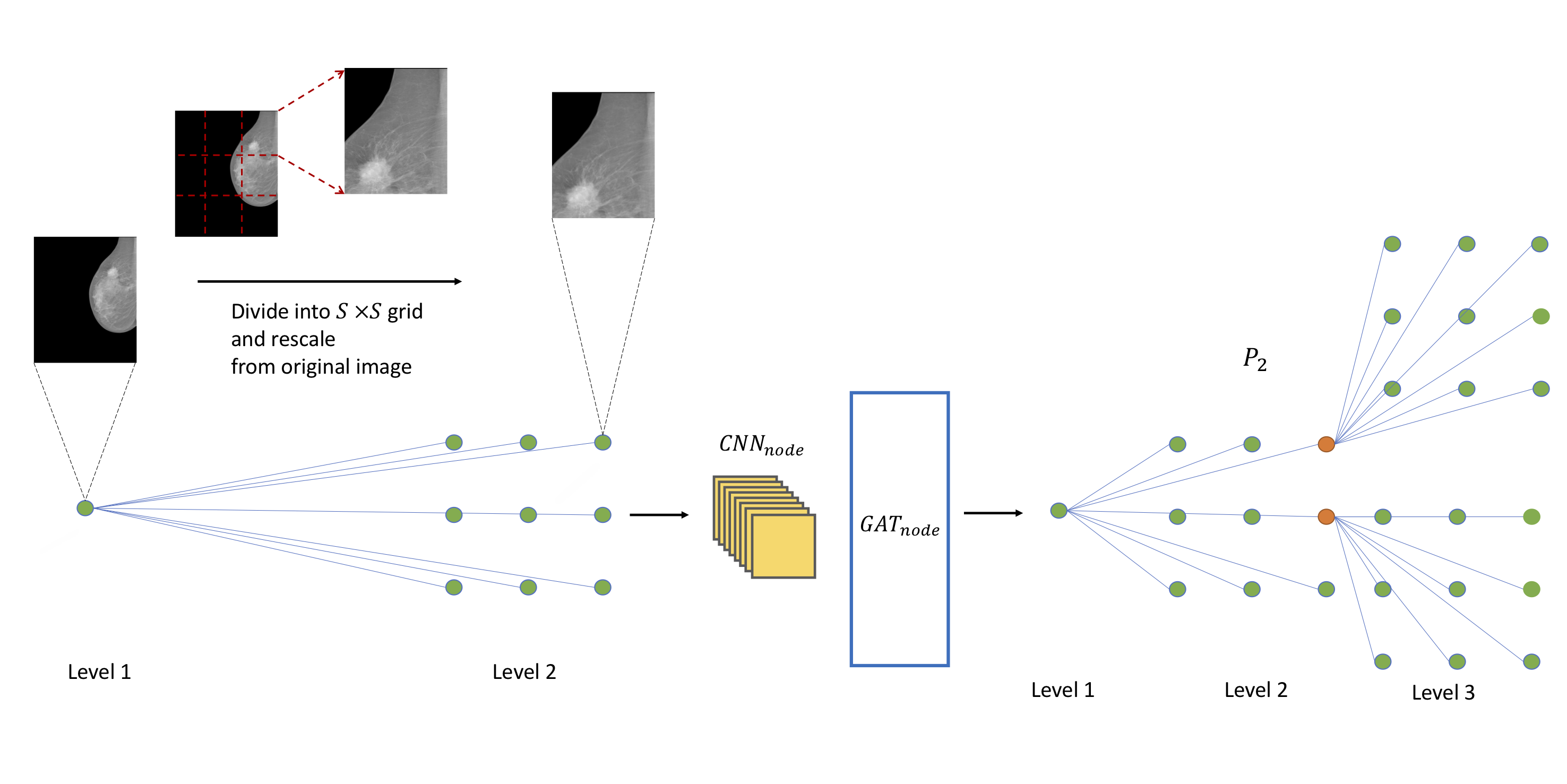} \\
		\caption{A demonstration of zoom-in mechanism in level $r$ = 2: each node in level $r$ represents a grid region in the connected node from previous level. $CNN_{node}$ and $GAT_{node}$ are applied to perform node classification. The nodes labelled in orange color are predicted to be zoomed in $r+1$ level. $G_r$ grows to $G_{r+1}$ accordingly.}
		\label{fig:sfig1}
	\end{subfigure}

	\begin{subfigure}{\textwidth}
		\centering
		\includegraphics[width=\textwidth]{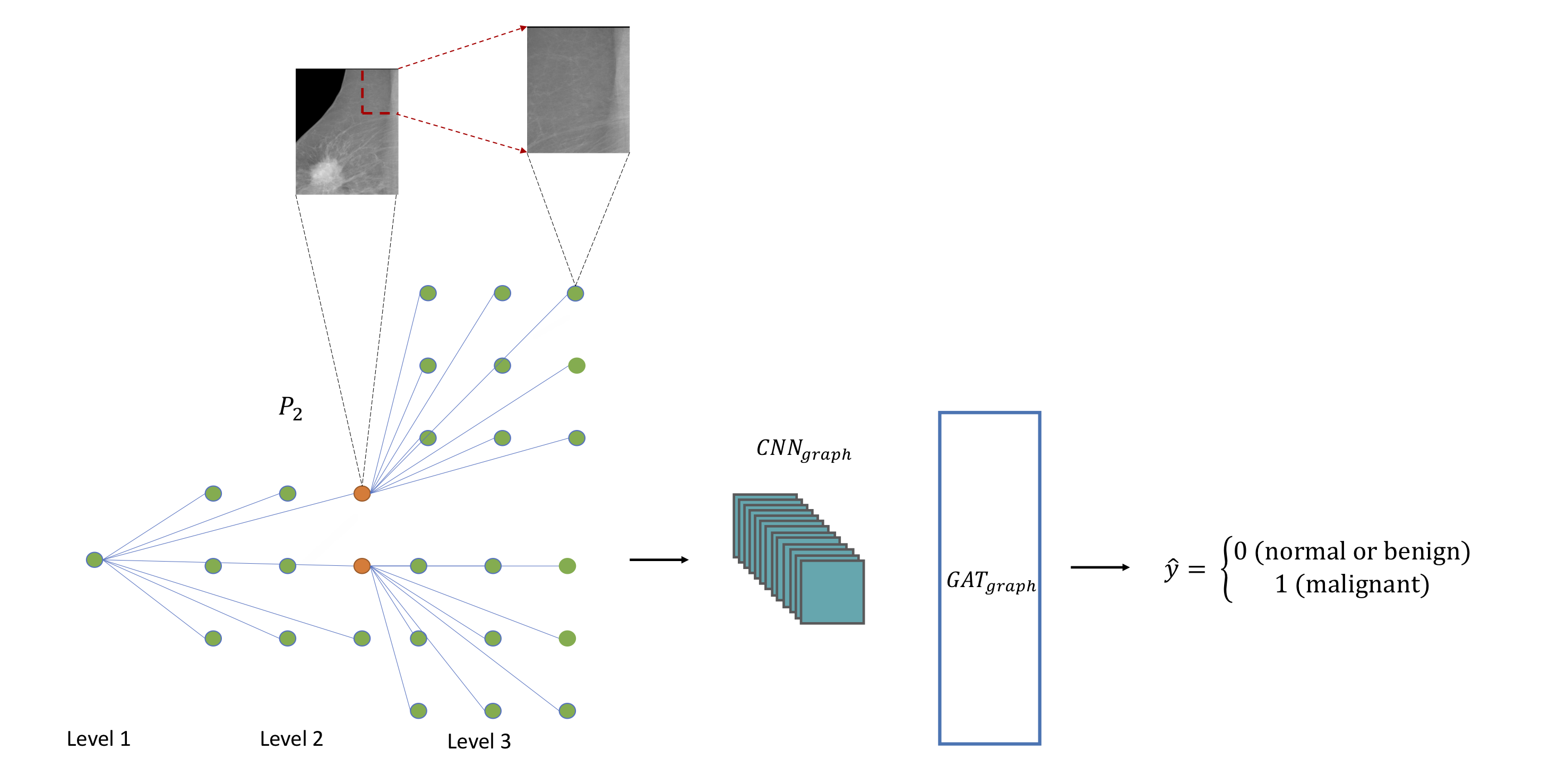}
		\caption{When $r = R$, full graph is fed into $CNN_{graph}$ and $GAT_{graph}$ for preform graph classification into binary classes.}
		\label{fig:sfig1}
	\end{subfigure}
	
\caption{An illustration of proposed network}
\label{fig:2}

\end{figure*}
\subsection{Problem Definition}
The mammography dataset is represented by $\mathcal{D} = \{(x^{(j)}, l^{(j)},y^{(j)})_{i}\}$ where $j \in \{left, right\}$, indicating the left and right breast of a patient in dataset. $x^{(j)} = \{x^{(j)}_{cc}, x^{(j)}_{mlo}\}$ represents the CC and MLO views of each side of a patient. $l^{(j)} = \{l^{(j)}_{cc}, l^{(j)}_{mlo}\}$ represents the segmentation labels of lesions (mass and calcification): $\Omega \rightarrow \{0, 1, 2, 3, 4\}$, where 0 indicates healthy or normal, 1 indicates benign calcification, 2 indicates benign mass, 3 indicates malignant calcification and 4 indicates malignant mass. $y \in \{0, 1\}$ represents the class label of a mammogram, where 0 indicates normal (BI-RADS $\in \{1, 2, 3\}$ and 1 indicates malignant (BI-RADS $\in \{4, 5, 6\}$. 

We use a graph instance to model the zooming operations in a mammogram $(x^{(j)})_i$, which is denoted as $(g^{(j)})_i = (V^{(j)}, A^{(j)}, X^{(j)})_i$. $(V^{(j)})_i$ is the node set and $|V^{(j)}_i| = n^{(j)}_i$, which indicates the number of zoomed regions in the mammogram. $A^{(j)}_i$ is an $n^{(j)}_i \times n^{(j)}_i$ matrix representing the connectivity in $(g^{(j)})_i$ and $(X^{(j)})_i \in n^{(j)}_i \times m \times m $ is a matrix recording the zoom-in regions of all nodes in $g$. The original mammogram dataset is transformed into a graph instance dataset $G = \{(g^{(j)}, l^{(j)}, y^{(j)})_{i}\}$. The zoom-in operation in this study refers that in a down-sized image, for a small region $M_{w \times h}$, we find the corresponding region in the original image and resize it to a larger image $M^\prime_{w^\prime \times h^{\prime}}$, where $w^\prime \times h^{\prime} > w \times h$. $M^{\prime}$ includes more information from original image compared to $M$.
\subsection{Network Design} 
In this paper, we study the mammogram classification as a graph classification problem. The graph instances and their connections are model as a hierarchical graph, as demonstrated in Figure 1. The number of zooming levels in the hierarchical graph is denoted as $R$. Top node A represents the full mammogram image. The levels after are created by zoom-in operations. Let us denote the graphs truncated at different zoom-in levels by $G^1, G^2, \dots, G^R$. At level $r$, graph is defined by its adjacency matrix $A_r \in \mathbb{R}^{N_r \times N_r}$. Its feature matrix is defined as $X_r \in \mathbb{R}^{N_r \times D \times D}$ (features on the nodes are cropped zoom-in regions from mammogram images, resized to $D \times D$). A convolutional neural network (CNN), pre-trained on lesion patches is used to extract feature vectors $H_r \in \mathbb{R}^{N_r \times H}$ from original features $X_r$. A graph attention network (GAT\textsubscript{node}) is used to classify node into two classes: to zoom into next level and not to zoom into next level. The output of $r$th level in the hierarchical graph is: 
\begin{equation}
\label{eq:1}
P_r = 
\begin{cases}
1 & r = 1 \\
\text{softmax}(GAT_{node}(A_r, CNN_{node}(X_r))) & 1 < r \leq R \\ 
\end{cases}
\end{equation}
The elements in $P_r \in \mathbb{R}^{N_r \times 2}$ gives the probability zooming into the next level in the hierarchical graph. For (i, j)th elements in $P_r$, if the prediction is to zoom into the next level, $S \times S$ nodes will be generated in the next level and mapped to (i, j)th node. The features on this node is divided into a $S \times S$ grid, resized to the same dimension $D \times D$ and assigned to the generated nodes in the next level. After the operation is performed on all nodes in $r$th level, there are $K$ nodes predicted to be zoomed into the next level. $G^{r+1}$ and $X^{r+1}$ will generated accordingly with $N_{r+1} \leftarrow N_r + K \times S^2$, $A_{r+1} \in \mathbb{R}^{N_{r+1} \times N_{r+1}}$ and $X_{r+1} \in \mathbb{R}^{N_{r+1} \times D \times D}$.

At the final zoom level $R$, an attention based GNN (GAT\textsubscript{graph}) is used to perform graph classification to classify the mammogram into normal/benign or malignant as follows: 
\begin{equation}
\label{eq:2}
\begin{aligned}
H^R_{graph} &= GAT_{graph}(A_R, CNN_{graph}(X_R)) \\
\hat{Y} &= \text{softmax}(H^R_{graph}W)
\end{aligned}
\end{equation}

The objective function for graph classification is defined as cross entropy loss between prediction and mammogram label: 
\begin{equation}
\label{eq:3}
\mathcal{L}_{graph} = \frac{1}{|\mathcal{D}|}\sum_{i, j}^{}y^{(j)}_i \log \hat{y}^{(j)}_i + (1 - y^{(j)}_i) \log (1 - \hat{y}^{(j)}_i)
\end{equation}

In addition to $\mathcal{L}_{graph}$, we construct node loss from zoom labels to supervise the zoom-in operation in node classification network: 
\begin{equation}
\label{eq:4}
\mathcal{L}_{node} = \frac{1}{|\mathcal{D}|} \frac{1}{N_R} \sum_{i, j, v}^{}z^{(j)}_{i, v} \log p^{(j)}_{i, v} + (1 - z^{(j)}_{i, v}) \log (1 - p^{(j)}_{i, v})
\end{equation}
Zoom labels are obtained from lesion segmentation label $(l^{(j)})_i$. For a node $v$ in final hierarchical graph $G_R$, the zoom label $z_v$ is obtained by from the lesion segmentation label of zoom-in region, $l^{(j)}_{i, v}$, corresponding to node $v$. We take the maximum label of $l^{(j)}_{i, v}$. If the maximum label is malignant, the zoom label $z^{(j)}_{i, v}$ is defined to 1, else 0: 
\begin{equation}
\label{eq:5}
z^{(j)}_{i, v} = 
\begin{cases}
1 &\text{if } max(l^{(j)}_{i, v}) \in \{0, 1, 2\} \\
0 &\text{if } max(l^{(j)}_{i, v}) \in \{3, 4\} \\
\end{cases}
\end{equation}

\section{Experiments and Performance Evaluation}

\textbf{Datasets and Baseline Algorithms: }The proposed method is evaluated using the INbreast \cite{moreira2012inbreast} dataset. INbreast is a publicly available  database with 116 cases comprising 410 FFDM images. INbreast dataset contains the BI-RADS assessment categories \cite{orel1999bi} on each mammogram. In addition, INbreast contains pixel-level lesion segmentation labels on each lesion detected by radiologists. The model performance is evaluated using area under receiver operating characteristic curve (AUC) \cite{huang2005using}. We compare the performance of the proposed model with state-of-the-art models on INbreast dataset, developed by \citeauthor{ribli2018detecting} and \citeauthor{shen2019deep} \cite{ribli2018detecting,shen2019deep}. 

\noindent\textbf{Experiment Setup: } The experiments are carried out with 80\% of the cases for training and 20\% for testing. The maximum number of zoom-in levels $R$ is selected to be 2. Zoom-in resize dimension $D$ is selected to be 224. The feature extraction CNN is defined as a VGG like network pre-trained on lesion patches, as demonstrated in \cite{nikulin2017digital,shen2019deep}. Zoom-in $S \times S$ grid is selected to be $3 \times 3$. The code and model of the experiment will be publicly available online for reproduction of this work.

%\begin{table}[]
%	\begin{tabular}{|l|l|}
%		\hline
%		Method & Testing AUC \\ \hline \hline
%		1      & 1           \\ \hline
%		2      & 2           \\ \hline
%		3      & 3           \\ \hline
%	\end{tabular}
%\label{tabel:1}
%\end{table}

\begin{figure}[ht!]
	\centering
	\includegraphics[width=0.5\textwidth]{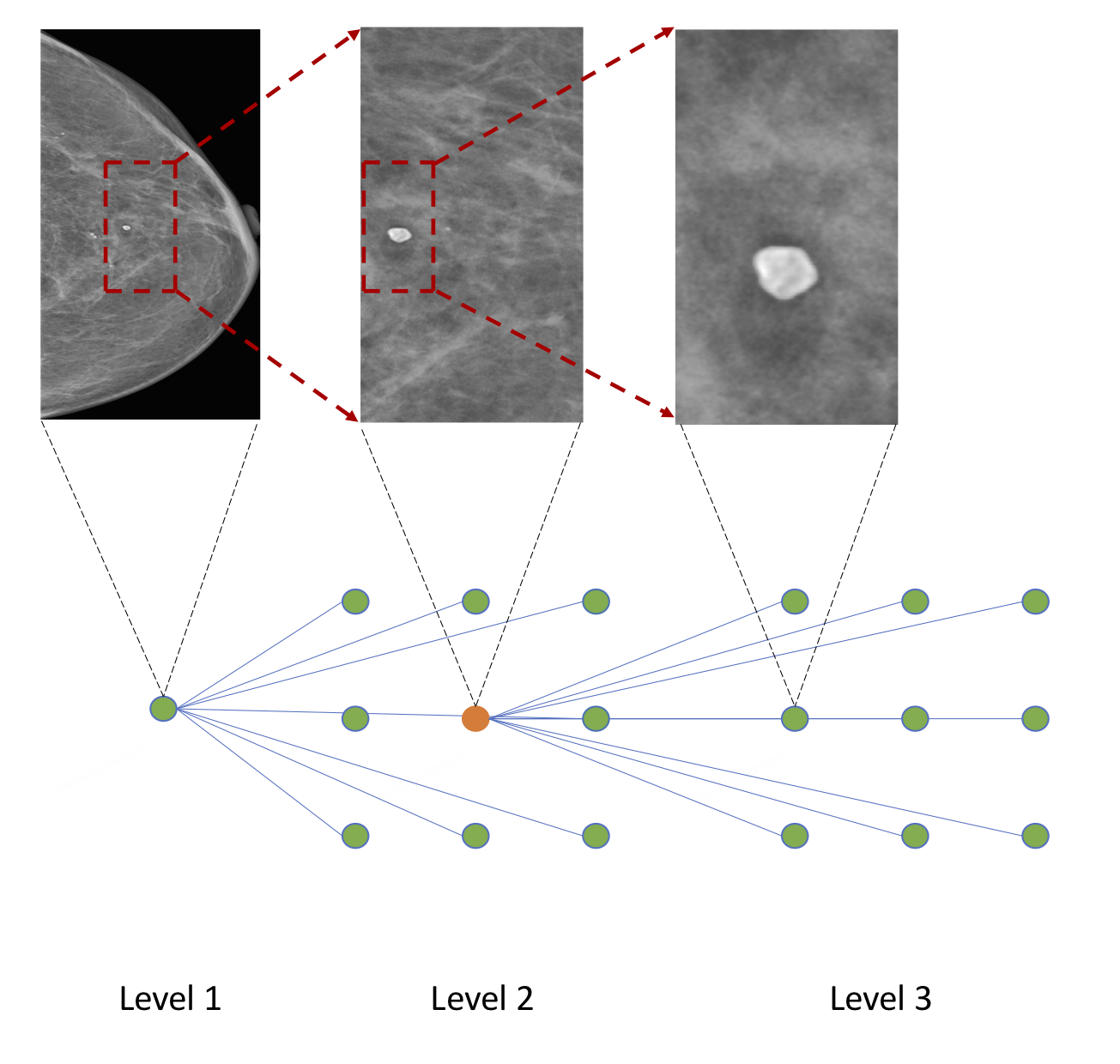} 
	\caption{An example data and its result from testing dataset. The images are features assigned to nodes accordingly. In level 2, the node in orange color is predicted to be zoomed in the next level and the corresponding region contains a cluster of malignant calcifications. }
	\label{fig:3}
	
\end{figure}

\noindent\textbf{Performance Analysis: } Our proposed method obtained AUC 0.943 in mammogram classification into normal/benign breast or malignant breast cancer. One example from testing set was selected for demonstration purpose in Figure \ref{fig:3}. In Figure \ref{fig:3}, there are several calcifications (tiny, bright, grey dots marked) in the mammogram and the proposed model is able to capture the calcification and successfully zoom into the specific region in the next two levels. Compare with the state-of-the-art models by \cite{shen2019deep} with AUC 0.95 and \cite{ribli2018detecting} with AUC 0.95, our proposed method produces similar results. The advantages of our model are: firstly, our model improves model interpretability by adding in zoom-in mechanism. The model can highlight the ROIs of lesions for clinicians both as a reminder in case they may have missed them or as a confirmation of their diagnosis. Secondly, our method collapses pixel-level lesion segmentation to zoom label. Compared to state-of-the-arts methods using pixel-level labels, our method is more robust to annotation errors. 
\\
\noindent\textbf{Limitations and future work: } There are a number of areas that we plan to continue working on to improve our model: firstly, the INbreast dataset is a relatively small dataset to demonstrate the capabilities of the method. Inspired by the achieved promising results, we plan to move on further validate the proposed algorithm locally at our own hospital with the ultimate goal to deploy it as an decision support tool for breast cancer screening. Secondly, the node classification network is difficult to be optimized using cross entropy loss on the zoom label. We plan to further investigate the problem and design a loss to better supervise the zoom-in mechanism in further studies. 
Lastly, we are now formally evaluating the robustness of our method against annotation noise in the actual clinical settings, and we will report our findings in our future work.

\section{Acknowledgements}
This research was supported by the National Research Foundation Singapore under its AI Singapore Programme [Award No. AISG-GC-2019-002] and Health Service Research Grant HSRG-OC17nov004.

~\\~
\\~
\\~
\\~
\\~
\\~
\\~
\\~
\\~
\\~
\\~
\\~
\\~
\\~
\\~
\\~
% \\~
% \\~
% \\~
% \\~
% \\~
% \\~
% \\~
% \\~

\pagebreak
\clearpage
\bibliography{aaaibib}
\bibliographystyle{aaai}
\end{document}